\documentclass[aps,pra,reprint,superscriptaddress,nofootinbib]{revtex4-2}

\usepackage{amsmath,amssymb,bm}
\usepackage{graphicx}
\usepackage{booktabs}
\usepackage{array}
\usepackage{xcolor}
\usepackage{hyperref}
\usepackage{enumitem}
\usepackage{tabularx}
\usepackage{placeins}
\graphicspath{{figures/}{figures_appendix/}{figures_notebook/}}
\hypersetup{colorlinks=true,linkcolor=blue!45!black,citecolor=blue!45!black,urlcolor=blue!45!black}
\setlist[itemize]{itemsep=1pt,topsep=2pt,leftmargin=1.2em}

\newcommand{\R}{\mathbb{R}}
\newcommand{\ket}[1]{|#1\rangle}

\begin{document}

\title{Physics-Informed Learning of Effective Error Processes from Limited Noisy Transmon Measurements for Robust QAOA Reliability}
\author{Ebrahim Khaleghian$^{1}$, Özgür~E.~Müstecaplıoğlu}
\affiliation{Department of Physics, Ko\c{c} University, Sarıyer, 34450, Istanbul, T\"urkiye}
\email{ebrahim.khaleghian@gmail.com}
\affiliation{TUBITAK Research Institute for Fundamental Sciences (TBAE), Gebze 41470, T\"urkiye}
\date{\today}

\begin{abstract}
We study whether limited finite-shot calibration measurements from a hidden transmon-inspired simulator can be used to learn compact, task-relevant effective error models for variational-algorithm reliability. Each physical element is modeled as a weakly anharmonic qutrit with coherent control imperfections, dissipation, dephasing, leakage, readout assignment error, and sampling noise, while the learner receives only selected local tomography data and, in the three-qubit extension, targeted pair probes. The learned objects are local affine Bloch response maps supplemented by process-relative edge residuals, and they are tested through their ability to reproduce and mitigate deformations of QAOA/MaxCut cost landscapes in simulation. The results show that strongly incomplete local data can still support useful response-map inference, that regularized linear models become competitive with neural networks in the scaled three-qubit setting, and that pair probes provide useful edge-level information, giving the clearest edge-residual prediction gain at the full pair-probe budget and a measurable downstream QAOA benefit. In the best non-oracle test, a Clifford-data-regression-style local-inverse correction reduces the finite-shot QAOA landscape error from $0.18187\,[0.17576,0.18798]$ to $0.01811\,[0.01751,0.01871]$, corresponding to a
$10\times$ improvement. The study supports a hardware-aware, measurement-efficient calibration strategy, while incorporating leakage-explicit diagnostics and a non-oracle
regression-style correction.

\end{abstract}

\maketitle

\section{Introduction}

Noisy intermediate-scale quantum processors execute algorithms through device-level dynamics rather than through idealized unitary gates.  In superconducting circuits, the measured response is shaped by weak transmon anharmonicity, coherent calibration errors, residual couplings, dissipation, dephasing, leakage outside the computational subspace, readout assignment error, and finite sampling noise~\cite{Koch2007Transmon,Krantz2019Guide,Blais2021CircuitQED,Motzoi2009DRAG}.  These mechanisms collectively deform the measured cost landscape and can therefore change the parameter values selected by a classical optimizer in a variational algorithm.

Quantum error mitigation (QEM) attempts to reduce such bias without full fault tolerance~\cite{Temme2017QEM,Endo2018PracticalQEM,Cai2023QEMReview}.  Standard mitigation strategies often require additional circuit executions, noise-scaling assumptions, quasi-probability overheads, or training circuits whose relation to the underlying device physics is only indirect.  In parallel, quantum process tomography, gate-set tomography, compressed tomography, and machine-learning-assisted tomography provide increasingly sophisticated routes to process characterization~\cite{Nielsen2021GST,Rudinger2021SimGST,Gross2010Compressed,Rodionov2014CompressedQPT,Torlai2023TensorQPT,Gaikwad2024NNTomography,Ahmed2023KrausLearning}. We investigate whether a small set of physically
motivated calibration measurements can learn a compact
response model accurate enough for algorithm-level error
mitigation.

The target benchmark is depth-one QAOA for MaxCut~\cite{Farhi2014QAOA,Zhou2020QAOA,Blekos2024QAOAReview}.  Although two- and three-qubit MaxCut instances are computationally easy, QAOA provides a controlled setting in which hardware-induced deformations of a cost landscape can be quantified. The hidden simulator contains qutrit transmon dynamics, Hamiltonian-level imperfections, Lindblad noise, leakage, readout assignment error, and finite-shot sampling. The learner is given selected local tomography data and
targeted pair-probe measurements, while the microscopic
Hamiltonian and noise parameters remain hidden.

We consider a compact set of learned operational response models. Local single-qubit behavior is represented by affine Bloch response maps that summarize the net influence of imperfect state preparation, control, leakage, and measurement on logical Pauli expectation values. Pair probes augment this local description through process-relative edge residuals that quantify deviations from a factorized product of local response maps. This local-plus-edge structure is naturally aligned with the MaxCut cost operator, which is itself a sum of two-qubit edge observables.

\begin{figure*}[t]
\centering
\includegraphics[width=1\textwidth]{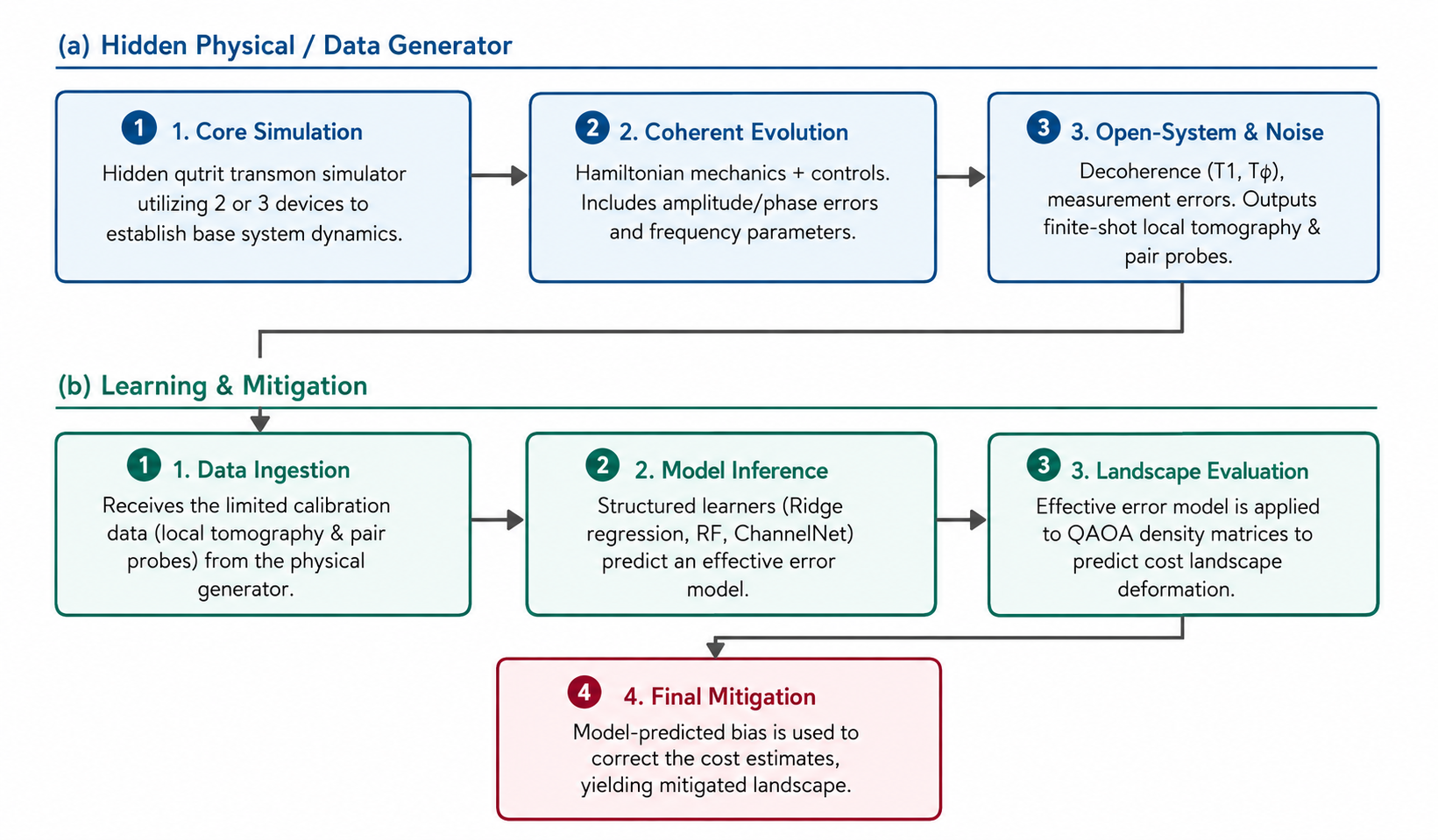}
\caption{Overall architecture. A hidden qutrit transmon simulator generates finite-shot
calibration data with coherent Hamiltonian imperfections, Lindblad noise,
leakage, readout assignment error, and sampling noise. The learner observes only
selected local tomography values and optional pair probes, then predicts compact
local response maps and process-relative edge residuals. QAOA landscapes are
evaluated at the logical-circuit level by applying the reference or learned
effective response model to ideal QAOA density matrices. Thus, the noisy QAOA
landscapes reported here are effective-response-model landscapes, not full
qutrit pulse-level simulations of the complete QAOA circuit.}
\label{fig:architecture}
\end{figure*}

The resulting workflow is summarized in Fig.~\ref{fig:architecture}. A hidden transmon-like simulator generates finite-shot calibration measurements, while the learning model receives only selected local tomography features and, when included, pair-probe data. From these measurements, it predicts compact local response maps and edge-level residuals, which are subsequently assessed by their ability to predict and mitigate distortions of the QAOA/MaxCut cost landscape.

The numerical study addresses three questions. First, can incomplete local calibration data support useful inference of compact response maps within a restricted but physically motivated family of transmon-like devices? Second, does the inverse problem require a nonlinear neural estimator, or can simpler structured models provide comparable or better performance? Third, do targeted pair probes provide useful edge-level observables beyond independent local descriptions, and does this additional information translate into improved QAOA reliability? We find that local response learning can substantially reduce landscape error, that Ridge regression is a strong and in the three-qubit setting often superior baseline, and that targeted pair probes improve process-relative edge-residual prediction with a measurable downstream QAOA benefit.

The measurement cost of the local-plus-edge protocol for an n-qubit device with edge set E scales as $O(n+|E|)$, or as $O(n)$ on bounded-degree hardware graphs, avoiding the exponential $O(16^n)$ parameter scaling associated with full tomography. This exponential contrast is the main resource-scaling motivation for learning compact response models.

\section{Hidden transmon physical model and noise taxonomy}

We model the hidden device as a network of weakly anharmonic transmons truncated to their three lowest energy levels. The simulator combines coherent Hamiltonian imperfections, open-system decoherence, leakage outside the computational subspace, readout assignment errors, and finite-shot sampling. These microscopic ingredients are used only to generate calibration measurements and reference response labels; they are not provided to the learning model. The learner instead receives selected noisy local tomography data and, when included, pair-probe measurements, from which it infers compact operational response models. The physical and measurement-level error mechanisms represented in the simulator are summarized in Appendix~\ref{app:implementation_details},
Fig.~\ref{fig:noise_taxonomy}. The QAOA benchmark is evaluated at the logical-circuit level. For each grid point, the ideal depth-one QAOA state is generated. The transmon-inspired qutrit simulator is used to generate finite-shot calibration data and reference effective response labels. The learned or reference response model is then applied to the ideal QAOA density matrix to predict the corresponding deformation of the cost landscape.

\subsection{Qutrit transmon truncation}

Each physical transmon is truncated to three levels, $\ket{0}$, $\ket{1}$, and $\ket{2}$.  The computational subspace is $\{\ket{0},\ket{1}\}$, while $\ket{2}$ allows leakage to be represented explicitly. This qutrit representation makes the hidden device more hardware-like than a purely two-level qubit model because control and coupling can populate noncomputational levels.

Let $a_i$ and $n_i=a_i^\dagger a_i$ denote the qutrit annihilation and number operators of transmon $i$, embedded in the full multi-transmon Hilbert space.  The simulator uses a time-dependent Hamiltonian
\begin{equation}
H(t)=H_0+H_{\rm ctrl}(t),
\end{equation}
where $H_0$ is the rotating-frame drift Hamiltonian and $H_{\rm ctrl}(t)$ represents the applied microwave control drives.  The drift Hamiltonian is
\begin{equation}
\begin{aligned}
H_0
=&\sum_i
\left[
\Delta_i n_i
-\frac{\alpha_i}{2}\,
n_i(n_i-1)
\right]  \\
&+\sum_{(i,j)}
g_{ij}
\left(
a_i^\dagger a_j
+
a_i a_j^\dagger
\right) \\
&+\sum_{(i,j)}
\zeta_{ij} n_i n_j .
\end{aligned}
\label{eq:h0}
\end{equation}
Here $\Delta_i$ represents detuning or calibration offset, $\alpha_i$ is the anharmonicity magnitude, $g_{ij}$ is exchange-like coupling, and $\zeta_{ij}$ is a residual $ZZ$-type interaction.

The control Hamiltonian is modeled through two microwave quadratures on each driven transmon,
\begin{equation}
H_{\rm ctrl}(t)
=
\sum_i
\left[
\frac{\widetilde{\Omega}_{x,i}(t)}{2}
\left(a_i+a_i^\dagger\right)
+
\frac{\widetilde{\Omega}_{y,i}(t)}{2}
i\left(a_i^\dagger-a_i\right)
\right].
\label{eq:hctrl}
\end{equation}
In the simulation, the control pulses are square pulses with nominal Rabi scale $\Omega=\theta/T$.  Hidden amplitude and phase errors transform the intended drive into an effective drive with scale $s_{\Omega,i}\Omega$ and phase $\phi_{\rm nominal}+\phi_i$.  The two quadratures implement rotations in the logical qubit subspace, while the qutrit operators allow the same drive to weakly populate the noncomputational level $\ket{2}$.

Thus $H_0$ describes always-present device physics such as detuning, anharmonicity, exchange coupling, and residual $ZZ$, while $H_{\rm ctrl}(t)$ describes the driven operations used for state preparation, tomography rotations. Both drift and control imperfections contribute to coherent Hamiltonian errors, and the full Hamiltonian $H(t)$ enters the Lindblad evolution below.

\subsection{Open-system dynamics}

The simulator includes Markovian open-system noise through a Lindblad master equation \cite{Lindblad1976,Gorini1976},
\begin{equation}
\dot{\rho}=-i[H,\rho]+\sum_k\left(L_k\rho L_k^\dagger-\frac{1}{2}\{L_k^\dagger L_k,\rho\}\right).
\label{eq:lindblad}
\end{equation}
The main collapse operators are
\begin{equation}
L_{1,i}=\sqrt{1/T_{1,i}}\,a_i,\qquad
L_{\phi,i}=\sqrt{1/T_{\phi,i}^{\rm coll}}\,n_i,
\end{equation}
representing energy relaxation and pure dephasing.  With the convention written here, the parameter $T_{\phi,i}^{\rm coll}$ is the collapse-operator parameter. In the three-qubit study, correlated dephasing is also included as a stress mechanism through a collective number operator. Relaxation changes populations; dephasing suppresses phase coherence; correlated dephasing produces errors that cannot always be attributed to independent local channels.

\subsection{Leakage, readout, and finite shots}

Leakage makes a qubit-only effective channel approximate because the physical
state may temporarily or permanently leave the logical subspace. In the hidden
qutrit simulator, leakage is represented by population in the noncomputational
level $|2\rangle$. This effect is made explicit
by recording leakage diagnostics in addition to logical Pauli expectations. For
transmon $i$, the leakage observable is
\begin{equation}
P_{\rm leak}^{(i)} = |2\rangle_i\langle 2| ,
\end{equation}
embedded in the full tensor-product Hilbert space, with leakage probability
\begin{equation}
p_{\rm leak}^{(i)} =
{\rm Tr}\!\left[\rho P_{\rm leak}^{(i)}\right].
\end{equation}
These leakage labels supplement the logical affine response model and provide a
direct diagnostic for population outside the computational subspace.

Readout error is modeled separately from evolution noise. Assignment
probabilities $r_{01}$ and $r_{10}$ represent misclassification of $|0\rangle$
as $|1\rangle$ and $|1\rangle$ as $|0\rangle$. Logical Pauli observables are
embedded in the qutrit space with zero action on the leakage level $|2\rangle$;
therefore, leaked population reduces the measured logical Pauli expectation.
Finite-shot measurement noise is introduced by sampling binary measurement
outcomes. For an observable with true expectation $m$, the observed estimate is
\begin{equation}
\hat m =
\frac{n_+ - n_-}{N_{\rm shots}},
\qquad
n_+ \sim
{\rm Binomial}\!\left(
N_{\rm shots},
\frac{1+m}{2}
\right).
\end{equation}
Thus Hamiltonian and Lindblad terms corrupt the quantum state, leakage is
tracked explicitly through $p_{\rm leak}^{(i)}$, and readout plus finite-shot
noise corrupt the observed data used for learning.

\section{Measurement protocols}

The measurement protocols define what information is visible to the learner.  The hidden simulator contains microscopic information such as detunings, couplings, decoherence rates, leakage, and readout errors. The learner is restricted to finite-shot estimates of selected Pauli expectation values. We test how accurately these limited measurement features determine compact effective response models.

\subsection{Local tomography}

For each qubit, the local tomography protocol prepares six logical input states,
\begin{equation}
\ket{0},\;\ket{1},\;\ket{+},\;\ket{-},\;\ket{+i},\;\ket{-i},
\end{equation}
and measures $X$, $Y$, and $Z$.  The states $\ket{0}$ and $\ket{1}$ probe the $Z$ axis of the Bloch sphere, $\ket{+}$ and $\ket{-}$ probe the $X$ axis, and $\ket{+i}$ and $\ket{-i}$ probe the $Y$ axis.  Measuring $X$, $Y$, and $Z$ after the noisy evolution provides information about how the device maps input Bloch vectors to output Bloch vectors.

Full local tomography therefore has
\begin{equation}
N_q\times 6\times 3
\end{equation}
finite-shot features: 36 for two qubits and 54 for three qubits.  These features are estimates of output Pauli expectations, not exact noiseless values.  The learner also receives $\log_{10}N_{\rm shots}$ as a shot-count feature so that it can distinguish low-shot and high-shot measurement regimes.

Partial tomography is implemented by masking a subset of local tomography entries.  If only $K$ settings are observed, the learner sees only those $K$ finite-shot features and must infer the remaining structure statistically.  In the two-qubit proof of concept, only 12 local tomography values are provided, while the target remains a 24-parameter local error model.  In the three-qubit setting, random local masks with $K=6,12,18,24,36,54$ are used.  The parameter $K$ therefore represents the local measurement budget: increasing $K$ gives the learner more direct information but also corresponds to more experimental measurement effort.

\subsection{Pair probes}

Local tomography only measures one-qubit responses.  It can learn how each qubit is distorted individually, but it may miss errors that appear only when two qubits are considered together.  For the three-qubit chain, pair probes measure products
\begin{equation}
\langle \sigma_i^a\sigma_j^b\rangle,\qquad a,b\in\{X,Y,Z\},
\end{equation}
for edges $(1,2)$ and $(2,3)$.  This gives $2\times 9=18$ pair-probe settings.

Pair probes target edge-level correlations that are poorly constrained by local tomography. Residual ZZ coupling, shared dephasing, and crosstalk can change two-qubit Pauli correlations even when local one-qubit expectations remain close to their expected values. Pair probes therefore provide direct information about edge-level error structure, consistent with the edge-based form of the MaxCut cost.

\section{Effective error-process representations}

Here we use compact effective representations rather than full multiqubit process matrices.  A full process description grows rapidly with the number of qubits and is unnecessary for the present goal.  The goal is to learn the parts of the error process that are most useful for predicting and mitigating the QAOA/MaxCut cost landscape. We therefore evaluate both response-map accuracy and QAOA landscape reliability. A model can estimate channel parameters well but still fail to correct the specific observable used by an algorithm. Our final metric is the reduction of QAOA/MaxCut landscape error, supplemented by channel MSE and L2 error. Here ``reliability'' means closeness of the noisy or mitigated algorithmic output to the ideal noiseless output.

\subsection{Local affine Bloch channels}

The local effective channel for qubit $i$ is represented as an affine map on the Bloch vector,
\begin{equation}
\bm r_{\rm out}^{(i)} = A_i \bm r_{\rm in}^{(i)} + \bm b_i,
\label{eq:affine_channel}
\end{equation}
where $A_i\in\R^{3\times 3}$ and $\bm b_i\in\R^3$. Each qubit therefore contributes 12 real parameters: nine from the matrix $A_i$ and three from the shift vector $\bm b_i$.  The two-qubit model has 24 parameters and the three-qubit model has 36.

The entries of $A_i$ and $\bm b_i$ have a direct operational interpretation in the affine Bloch representation of a qubit response map. With leakage, readout error, and finite-shot sampling, the affine map is an operational response representation rather than a microscopic CPTP channel. In the leakage-free and SPAM-free limit it reduces to the usual affine representation of a qubit quantum channel \cite{NielsenChuang,KingRuskai2001,Ruskai2002}. 
Diagonal contractions in $A_i$ describe reductions of Bloch-vector length, off-diagonal components describe rotations or mixing between Bloch-sphere axes, and $b_i$ characterizes nonunital shifts such as amplitude-damping-type relaxation. In the present qutrit simulator, this map represents the projected logical response after qutrit evolution, leakage, readout assignment, and finite-shot estimation \cite{WoodGambetta2018}.

\subsection{Process-relative pairwise residuals}

Local affine channels assume that each qubit can be modeled independently. Multiqubit hardware also produces edge-level response components from residual (ZZ) coupling, shared
dephasing, crosstalk, and leakage that cannot be represented as a tensor product of independent local maps. We
therefore supplement the local affine model with process-relative pairwise
residuals.

For an edge $(i,j)$, let $P_{\rm in}^{ab}$ denote a two-qubit Pauli-pair probe,
with $a,b\in\{X,Y,Z\}$, and let $\sigma_i^c\sigma_j^d$ denote a two-qubit Pauli
measurement, with $c,d\in\{X,Y,Z\}$. The hidden simulator defines the noisy edge
response
\begin{equation}
R_{ij}^{ab\rightarrow cd}
=
\left\langle \sigma_i^c\sigma_j^d \right\rangle_{\rm true}
\end{equation}
for that probe. The product-local model predicts the corresponding response by
applying the two learned local response maps independently to the same probe
input,
\begin{equation}
R_{ij,{\rm local}}^{ab\rightarrow cd}
=
\left\langle \sigma_i^c \right\rangle_{\rm local}
\left\langle \sigma_j^d \right\rangle_{\rm local}.
\end{equation}
The process-relative pair residual is then defined as
\begin{equation}
\delta_{ij}^{ab\rightarrow cd}
=
R_{ij}^{ab\rightarrow cd}
-
R_{ij,{\rm local}}^{ab\rightarrow cd}.
\end{equation}
This definition compares the noisy edge response with the tensor product of the
corresponding local responses on the same calibration probe ensemble. For the three-qubit chain, the physical edges are $(1,2)$ and $(2,3)$. Each edge
has nine Pauli-pair input probes and nine Pauli-pair output responses, giving $2\times 9\times 9 = 162$ process-relative pair-residual labels.

\section{QAOA/MaxCut benchmark and reliability}

The target algorithm used to evaluate the learned error models is the Quantum Approximate Optimization Algorithm (QAOA) for MaxCut. QAOA is a hybrid variational quantum algorithm for combinatorial optimization: it prepares a parameterized quantum state, measures an objective function, and uses a classical optimizer or parameter scan to choose parameters that improve the measured objective value \cite{Farhi2014QAOA,Moll2018Variational,Zhou2020QAOA}. We use QAOA/MaxCut as a controlled algorithm-level
benchmark for testing response-model-based mitigation.

\subsection{MaxCut objective}

MaxCut is a graph partitioning problem.  Given a graph $G=(V,E)$, the task is to divide the vertices into two groups so that as many edges as possible connect vertices in different groups.  A classical bit string assigns each vertex to one of the two groups \cite{GoemansWilliamson1995,Farhi2014QAOA}.  It is convenient to represent the group assignment by variables $z_i=\pm 1$.  For one edge $(i,j)$, the edge is cut when the endpoints have opposite signs, $z_iz_j=-1$, and it is not cut when they have the same sign, $z_iz_j=+1$.  Therefore the classical edge contribution can be written as
\begin{equation}
C_{ij}(z)=\frac{1}{2}\left(1-z_i z_j\right).
\end{equation}
This equals 1 for a cut edge and 0 for an uncut edge.

The corresponding quantum cost operator is obtained by replacing the classical variables with Pauli-$Z$ operators,
\begin{equation}
C=\sum_{(i,j)\in E}\frac{1}{2}\left(I-Z_iZ_j\right).
\label{eq:maxcut_cost}
\end{equation}
Measuring this operator gives the expected MaxCut value of the quantum state. The two-qubit single-edge instance isolates the response-learning task: inferring a 24-parameter effective local response map from 12 noisy tomography values and using it to correct the QAOA landscape. The three-qubit study uses a chain with edges $(1,2)$ and $(2,3)$, matching the physical coupling topology and introducing a more meaningful scaling step.

\subsection{QAOA state and cost landscape}

For depth $p=1$, the ideal QAOA state is
\begin{equation}
\ket{\psi(\gamma,\beta)}
=
e^{-i\beta H_M}
e^{-i\gamma C}
\ket{+}^{\otimes n},
\label{eq:qaoa_state}
\end{equation}
where
\begin{equation}
H_M=\sum_i X_i
\end{equation}
is the mixer Hamiltonian.  The parameter $\gamma$ controls the problem-dependent cost evolution, while $\beta$ controls the mixing evolution.  The initial state $\ket{+}^{\otimes n}$ is an equal superposition over all computational-basis bit strings.

For each pair of parameters $(\gamma,\beta)$, the ideal expected cost is
\begin{equation}
C_{\rm ideal}(\gamma,\beta)
=
\langle \psi(\gamma,\beta)|C|\psi(\gamma,\beta)\rangle .
\end{equation}
Evaluating this expectation value over a grid of $\gamma$ and $\beta$ produces the QAOA cost landscape. Noise deforms this landscape and can shift the variational parameters selected by an optimizer.

A useful response model must predict the algorithmic
observable, not only the reference response labels. If noise changes the cost landscape, a classical optimizer may select parameters that look good on the noisy device but are suboptimal for the ideal objective.  The QAOA landscape is therefore used here as an algorithm-level validation test of the learned response model.

\section{Learning models, norms, and evaluation metrics}

The learning problem is supervised.  Each training example corresponds to one hidden device instance and one measurement protocol.  The input is a finite-shot tomography feature vector.  The target label is the effective response representation computed from the hidden simulator: local affine-map parameters and, in the pair-probe study, process-relative edge residuals.  At test time, the model receives only the measurement features and must predict these effective labels.

We compare direct reconstruction, a mean baseline, Ridge regression, Random Forest regression, ChannelNet, and a pair-probe-aware neural model.  The mean baseline ignores the measurement input and predicts an average label learned from the training set.  Direct reconstruction uses the measured tomography entries directly where possible, but becomes weak when the tomography data are incomplete.  Ridge regression is a regularized linear model and tests whether the inverse map is approximately linear.  Random Forest regression provides a nonlinear non-neural baseline.  ChannelNet is a multilayer neural network designed to learn nonlinear structure in the mapping from limited measurements to effective response parameters.  The pair-probe-aware model extends this idea by predicting both local response maps and process-relative edge residuals.

This comparison tests whether the inverse map from limited measurements to effective response parameters is adequately captured by linear regularization or requires nonlinear estimators. Including Ridge and Random Forest baselines follows current ML-QEM practice, where simple models can be competitive with neural networks \cite{Liao2024MLQEM}.

The channel-label mean-squared error is
\begin{equation}
\mathrm{MSE}=\frac{1}{d}\|\hat{\bm y}-\bm y\|_2^2,
\end{equation}
where $\bm y$ is the flattened channel-label vector, $\hat{\bm y}$ is the predicted label vector, and $d$ is the number of label components.  MSE measures the average squared error per channel parameter.  The per-device L2 error is
\begin{equation}
\mathrm{L2}^{(s)}=\|\hat{\bm y}^{(s)}-\bm y^{(s)}\|_2.
\end{equation}
This measures the total channel-label error for one test device $s$.  Reporting both MSE and L2 is useful because MSE gives a normalized component-wise error, while L2 gives the total size of the prediction error per device.

For pair residuals, the same L2 norm is reported together with a mean absolute pair error,
\begin{equation}
\mathrm{MAE}_{\rm pair}=\frac{1}{d_{\rm pair}}\sum_k |\hat\delta_k-\delta_k|.
\end{equation}
The pair L2 error measures the total edge-residual prediction error, while $\mathrm{MAE}_{\rm pair}$ gives the average absolute error per residual component.

QAOA reliability is evaluated through landscape-level metrics.  The main metric is the mean absolute error between an estimated or mitigated cost landscape and the ideal landscape,
\begin{equation}
\mathrm{MAE}_{\rm QAOA}=\frac{1}{|\Gamma||B|}\sum_{\gamma\in\Gamma,\beta\in B}|C_{\rm est}(\gamma,\beta)-C_{\rm ideal}(\gamma,\beta)|.
\end{equation}
Here $\Gamma$ and $B$ are the grids of QAOA parameters.  A smaller $\mathrm{MAE}_{\rm QAOA}$ means that the noisy or corrected landscape is closer to the ideal noiseless landscape.

The improvement ratio is
\begin{equation}
\mathcal{I}=\frac{\mathrm{MAE}_{\rm noisy}}{\mathrm{MAE}_{\rm mitigated}}.
\end{equation}

We also report RMSE, regret, approximation ratio, optimal-bitstring probability $P_{\rm opt}$, and parameter displacement. These quantities separate landscape-level accuracy from parameter-selection quality and final sampling success \cite{WillmottMatsuura2005,Farhi2014QAOA,Zhou2020QAOA,Blekos2024QAOAReview}.

We use two related mitigation evaluations. The first is an oracle-assisted
simulation diagnostic. From a learned effective error model, we compute a
model-predicted noisy landscape $C_{\rm model}(\gamma,\beta)$. Since the
noiseless landscape is available in simulation, the model-predicted bias can be
defined as
\begin{equation}
\hat b_C(\gamma,\beta)
=
C_{\rm model}(\gamma,\beta)-C_{\rm ideal}(\gamma,\beta).
\end{equation}
The oracle-diagnostic corrected landscape is then
\begin{equation}
C_{\rm mit}^{\rm oracle}(\gamma,\beta)
=
C_{\rm noisy}(\gamma,\beta)-\hat b_C(\gamma,\beta).
\end{equation}
Here $C_{\rm noisy}$ denotes the finite-shot landscape obtained by applying the reference effective response model to ideal logical QAOA states. This oracle-assisted diagnostic tests whether the learned response model captures the simulated landscape deformation.

To obtain a non-oracle mitigation test, we also implement a
Clifford-data-regression-style correction. A set of auxiliary classically
tractable training circuits is generated from coarse QAOA-like parameter grids.
For each training circuit $t$, the noisy finite-shot observable features \(\phi^{(t)}\) are paired with the exactly computable ideal value
$C_{\rm ideal}^{(t)}$. A regression map is then trained to predict the ideal
value from the noisy or locally corrected observable features,
\begin{equation}
f_{\rm CDR}:\; \phi^{(t)}
\longmapsto
C_{\rm ideal}^{(t)} .
\end{equation}
At test time, the learned map is applied directly to the target noisy QAOA
landscape,
\begin{equation}
C_{\rm mit}^{\rm CDR}(\gamma,\beta)
=
f_{\rm CDR}\!\left(\phi(\gamma,\beta)\right).
\end{equation}
In the local-inverse variant, a locally inverse-corrected cost estimate,
computed from the learned local response maps, is included as an additional
regression feature. The CDR local-inverse correction uses ideal values only for auxiliary classically tractable calibration circuits; the target ideal landscape is not inserted.
The CDR-style local-inverse correction is the non-oracle mitigation protocol tested here. The response-model stage learns compact effective errors from tomography and pair-probe features; the CDR layer then maps noisy or locally corrected observables to ideal values using auxiliary training circuits \cite{Strikis2021LearningQEM,Czarnik2021CDR}.

\begin{figure*}[t]
\centering
\begin{minipage}{0.45\textwidth}
\centering
\includegraphics[width=\linewidth]{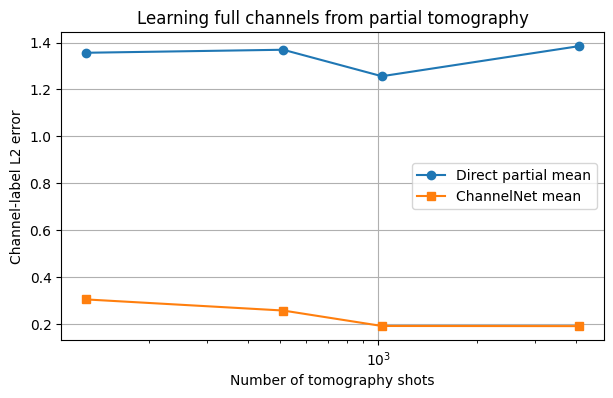}\\[-1mm]
{\small (a) Partial-tomography channel inference.}
\end{minipage}\hfill
\begin{minipage}{0.45\textwidth}
\centering
\includegraphics[width=\linewidth]{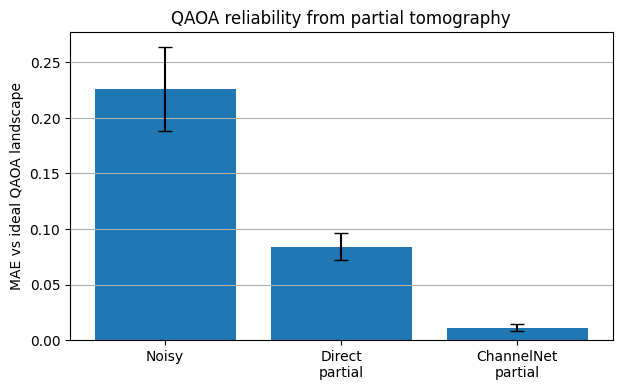}\\[-1mm]
{\small (b) QAOA reliability over hidden devices.}
\end{minipage}
\caption{Two-qubit case. In the underdetermined partial-tomography setting, ChannelNet learns a structured prior over hardware-generated effective channels and substantially outperforms direct partial reconstruction.}
\label{fig:twoqubit_poc}
\end{figure*}

\section{Results I: two-qubit proof of concept}

The two-qubit experiment provides a controlled proof of concept. The single-edge MaxCut instance isolates the response-learning problem: inferring an effective hardware-induced distortion from limited noisy tomography and using it to correct the QAOA landscape. Complete tomography is used as a control case, where direct reconstruction is strong when all local measurements are available.  The main test is partial tomography.  The input contains only 12 local tomography values plus the shot feature, but the target is a 24-parameter local channel.  This is intentionally underdetermined for direct reconstruction.

Table~\ref{tab:twq_partial} and Fig.~\ref{fig:twoqubit_poc} summarize the two-qubit partial result.  ChannelNet reduces the channel MSE from $7.688\times 10^{-2}$ for direct partial tomography to $2.714\times10^{-3}$.  More importantly, it reduces QAOA landscape MAE from $0.22611$ for the noisy baseline to $0.01108$, corresponding to a $20.41\times$ improvement.  Direct partial tomography helps, but only reaches $0.08413$.

\begin{table}[t]
\caption{Two-qubit partial-tomography proof of concept.  Only 12 tomography values are used as input, while the target remains the full 24-parameter local channel.}
\label{tab:twq_partial}
\begin{ruledtabular}
\begin{tabular}{lccc}
Method & Channel MSE & Mean L2 & QAOA MAE \\
\hline
Noisy/mean & $1.058\times10^{-1}$ & 1.56282 & 0.22611 \\
Direct partial & $7.688\times10^{-2}$ & 1.34910 & 0.08413 \\
ChannelNet partial & $2.714\times10^{-3}$ & 0.23296 & 0.01108 \\
\end{tabular}
\end{ruledtabular}
\end{table}

\begin{figure*}[t]
\centering
\begin{minipage}{0.45\textwidth}
\centering
\includegraphics[width=\linewidth]{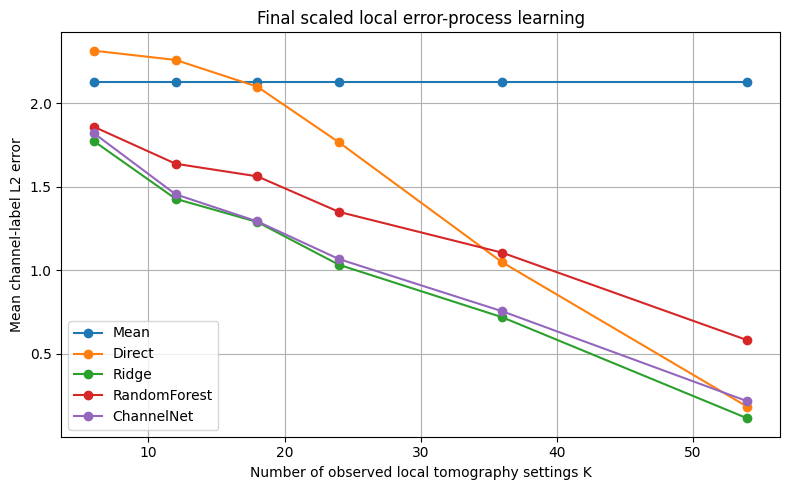}\\[-1mm]
{\small (a) Mean local-channel L2 error vs. $K$.}
\end{minipage}\hfill
\begin{minipage}{0.45\textwidth}
\centering
\includegraphics[width=\linewidth]{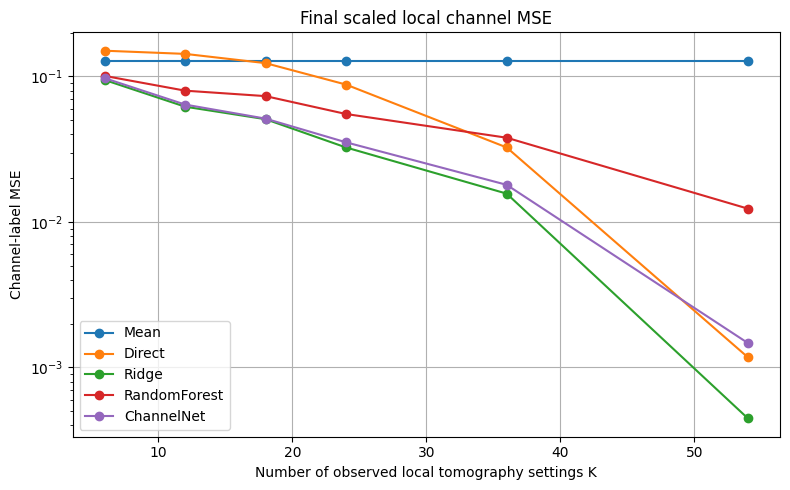}\\[-1mm]
{\small (b) Channel-label MSE vs. $K$.}
\end{minipage}
\caption{Three-qubit local error learning. The local-channel result is shown across the full observed-tomography sweep $K=6,12,18,24,36,54$. Ridge is the strongest local estimator in the scaled run, while ChannelNet remains competitive.}
\label{fig:threequbit_local}
\end{figure*}

\begin{table*}[t]
\centering
\caption{
Three-qubit local QAOA reliability at $K=18$ and $K=54$ using the
oracle-diagnostic local-response correction.  The noisy baseline is evaluated
separately at each budget because the finite-shot samples differ.  The MAE
confidence intervals are $95\%$ confidence intervals over $n=50$ hidden-device
test samples.
}
\begin{tabular}{clccccc}
\hline\hline
$K$ & Method & Mean MAE (95\% CI) & Mean RMSE & Mean regret & Mean $P_{\rm opt}$ & Improvement \\
\hline
18 & Noisy        & $0.177493\,[0.16491,0.19008]$ & $0.219523$ & $0.178446$ & $0.531209$ & $1.00\times$ \\
18 & Direct       & $0.138421\,[0.12553,0.15131]$ & $0.169831$ & $0.038852$ & $0.624388$ & $1.28\times$ \\
18 & Ridge        & $0.026895\,[0.02271,0.03109]$ & $0.032986$ & $0.008966$ & $0.661760$ & $6.60\times$ \\
18 & Random Forest & $0.039266\,[0.03219,0.04634]$ & $0.047647$ & $0.012080$ & $0.659320$ & $4.52\times$ \\
18 & ChannelNet   & $0.030598\,[0.02555,0.03565]$ & $0.037296$ & $0.009841$ & $0.659905$ & $5.80\times$ \\
\hline
54 & Noisy        & $0.177743\,[0.16520,0.19029]$ & $0.219710$ & $0.169638$ & $0.537988$ & $1.00\times$ \\
54 & Direct       & $0.019516\,[0.01685,0.02218]$ & $0.024054$ & $0.005043$ & $0.664296$ & $9.11\times$ \\
54 & Ridge        & $0.013445\,[0.01301,0.01388]$ & $0.016800$ & $0.005032$ & $0.665252$ & $13.22\times$ \\
54 & Random Forest & $0.022456\,[0.01747,0.02744]$ & $0.027589$ & $0.006605$ & $0.662345$ & $7.92\times$ \\
54 & ChannelNet   & $0.014271\,[0.01336,0.01518]$ & $0.017772$ & $0.005026$ & $0.665316$ & $12.45\times$ \\
\hline\hline
\end{tabular}
\label{tab:three_qubit_local_qaoa}
\end{table*}

\section{Results II: three-qubit local error learning}

The three-qubit local experiment tests whether the local-response learning idea
survives a harder setting. Full local tomography has 54 entries and the local
channel label has 36 parameters. Fig.~\ref{fig:threequbit_local} shows the final scaled local-channel
L2 error and channel-label MSE across the local tomography sweep
$K=6,12,18,24,36,54$. The QAOA reliability comparison is reported separately in
Table~II, where QAOA was evaluated at the representative budgets $K=18$ and
$K=54$.

At $K=18$, Ridge and ChannelNet have nearly identical channel L2 errors,
$1.2879$ and $1.2915$, respectively, and both substantially improve QAOA
reliability. Ridge reduces the QAOA landscape MAE from
$0.17749\,[0.16491,0.19008]$ for the noisy baseline to
$0.02690\,[0.02271,0.03109]$, corresponding to a $6.60\times$ improvement.
ChannelNet reduces the MAE to $0.03060\,[0.02555,0.03565]$, corresponding to a
$5.80\times$ improvement. At $K=54$, Ridge reaches a QAOA MAE of
$0.01345\,[0.01301,0.01388]$ and ChannelNet reaches
$0.01427\,[0.01336,0.01519]$, corresponding to improvement ratios of
$13.22\times$ and $12.45\times$.

The three-qubit setting leads to a different estimator hierarchy. In two qubits,
ChannelNet clearly dominates the underdetermined partial-reconstruction task.
In the scaled three-qubit local-channel task, Ridge is the strongest estimator.
This suggests that the mapping from randomly masked local tomography data to
local affine-channel labels is close to a regularized linear inverse problem
once enough measurements are available. A separate neural-network stability
check over three training seeds gives consistent ChannelNet QAOA performance.

Confidence intervals in Table~\ref{tab:three_qubit_local_qaoa} are computed
over hidden-device test samples; seed-to-seed neural-network stability is
reported separately.
\section{Results III: pair probes and correlated hardware errors}

Local tomography alone cannot fully identify correlated or edge-level response
components.  The three-qubit pair-probe experiment tests whether targeted pair
measurements improve the learning of nonlocal error structure.  The pair labels are process-relative residuals: for each edge,
the residual is defined as the difference between the true noisy pair response
and the response predicted by the tensor product of the learned local affine maps
on the same pair-probe ensemble.  For the three-qubit chain with edges $(1,2)$
and $(2,3)$, this gives $2\times 9\times 9=162$ pair-residual labels.

The pair-aware learner predicts the local affine channels together with these
process-relative pair residuals. Increasing the pair-probe budget improves downstream QAOA reliability monotonically and gives the clearest edge-residual prediction gain at the full
\(K_{\rm pair}=18\) budget. In the pair-aware QAOA
test, increasing $K_{\rm pair}$ from $0$ to $18$ reduces the QAOA landscape MAE
from $0.04958$ to $0.03535$. The corresponding improvement ratio increases from
$3.51\times$ to $4.93\times$. The added pair observables therefore provide useful edge-level information for QAOA landscape correction, although the component-wise residual-learning metrics are not strictly monotonic at the smallest pair-probe budget. Table III summarizes the process-relative residual-learning metrics, while Fig. 4 and Table IV summarize the corresponding pair-aware QAOA reliability results.

We also tested non-oracle ablations, including local-
only correction, pair-residual correction with shrinkage, and CDR-style correction. The shrinkage test is impor-
tant because noisy pair-residual estimates can overcorrect the target landscape.

\subsection{Non-oracle CDR-style mitigation}

We next evaluate a non-oracle CDR-style correction in which
ideal values enter only through auxiliary classically
tractable training circuits.

The best non-oracle method is the CDR-style local-inverse variant
with a coarse $7\times 7$ auxiliary training grid. It reduces the QAOA landscape
MAE from $0.18187\,[0.17576,0.18798]$ for the noisy edge-measured baseline to
$0.01811\,[0.01751,0.01871]$, corresponding to a $10.04\times$ improvement.
This is the main non-oracle mitigation result of the work; Tables~\ref{tab:three_qubit_local_qaoa} and \ref{tab:pair_qaoa_process_relative} report oracle-assisted diagnostics.

\begin{table*}[t]
\centering
\caption{
Three-qubit process-relative pair-probe learning. Pair probes provide additional edge-level observables beyond the product of local affine maps. The residual-learning metrics are not strictly monotonic at the smallest pair-probe budget, but the full \(K_{\rm pair}=18\) budget gives the clearest improvement in edge-residual prediction. The pair label contains $2\times 9\times 9=162$ process-relative residual components.}
\begin{ruledtabular}
\begin{tabular}{cccccc}
$K_{\rm pair}$ & Channel MSE & Mean channel L2 & Pair MSE & Mean pair L2 & Mean abs. pair error \\
\hline
0  & $4.9445\times 10^{-2}$ & $1.2793$ & $1.1875\times 10^{-1}$ & $4.3165$ & $0.2019$ \\
6  & $5.4763\times 10^{-2}$ & $1.3560$ & $1.2406\times 10^{-1}$ & $4.4308$ & $0.2074$ \\
12 & $5.1767\times 10^{-2}$ & $1.3083$ & $1.0257\times 10^{-1}$ & $3.9720$ & $0.1824$ \\
18 & $4.6662\times 10^{-2}$ & $1.2380$ & $9.2021\times 10^{-2}$ & $3.7581$ & $0.1722$
\label{tab:pair_probe_process_relative}
\end{tabular}
\end{ruledtabular}
\end{table*}

\begin{table*}[t]
\centering
\caption{
Three-qubit pair-aware QAOA reliability using process-relative pair residuals.
The noisy baseline is evaluated on the same hidden-device test set.  Increasing
the pair-probe budget improves the corrected QAOA landscape.
}
\begin{ruledtabular}
\begin{tabular}{cccccc}

Method & Mean MAE & Mean RMSE & Mean regret & Improvement \\
\hline
Noisy & $0.17422$ & $0.20868$ & $0.16716$ & $1.00\times$ \\
$K_{\rm pair}=0$ local+pair  & $0.04958$ & $0.05771$ & $0.01053$ & $3.51\times$ \\
$K_{\rm pair}=6$ local+pair  & $0.04609$ & $0.05417$ & $0.01122$ & $3.78\times$ \\
$K_{\rm pair}=12$ local+pair & $0.04214$ & $0.04969$ & $0.00965$ & $4.13\times$ \\
$K_{\rm pair}=18$ local+pair & $0.03535$ & $0.04172$ & $0.00646$ & $4.93\times$ \\
Oracle true local+pair       & $0.01213$ & $0.01524$ & $0.00598$ & $14.37\times$

\label{tab:pair_qaoa_process_relative}
\end{tabular}
\end{ruledtabular}
\end{table*}

\begin{figure*}[t]
\centering
\begin{minipage}{0.45\textwidth}
\centering
\includegraphics[width=\linewidth]{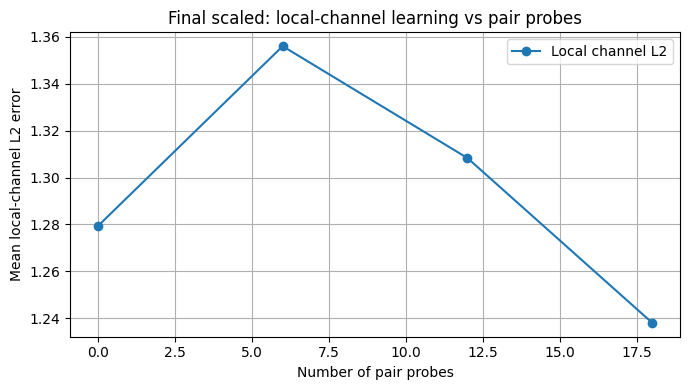}\\[-1mm]
{\small (a) Local-channel L2 vs. pair probes.}
\end{minipage}\hfill
\begin{minipage}{0.45\textwidth}
\centering
\includegraphics[width=\linewidth]{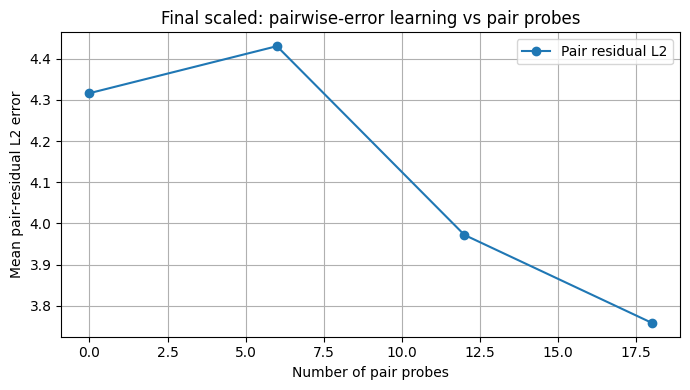}\\[-1mm]
{\small (b) Pair-residual L2 vs. pair probes.}
\end{minipage}
\vspace{1mm}
\begin{minipage}{0.45\textwidth}
\centering
\includegraphics[width=\linewidth]{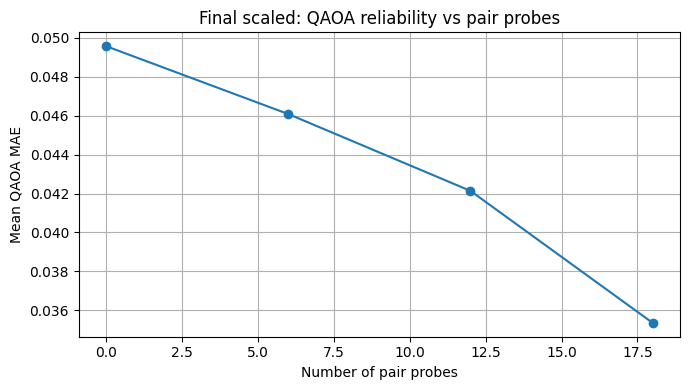}\\[-1mm]
{\small (c) QAOA MAE vs. pair probes.}
\end{minipage}\hfill
\begin{minipage}{0.45\textwidth}
\centering
\includegraphics[width=\linewidth]{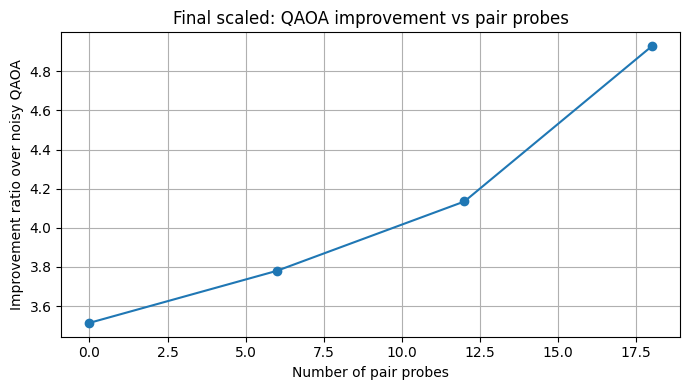}\\[-1mm]
{\small (d) QAOA improvement vs. pair probes.}
\end{minipage}
\caption{Pair probes provide useful edge-level information for process-relative edge-residual learning and downstream QAOA reliability. The residual labels compare the true noisy pair response with the tensor product of the learned local affine responses on the same pair-probe ensemble. Increasing the pair-probe budget reduces the pair-aware QAOA landscape error
monotonically, while the full \(K_{\rm pair}=18\) budget gives the clearest
edge-residual prediction gain.}
\label{fig:pair_probes}
\end{figure*}
\FloatBarrier
\section{Conclusion}

We have studied a physics-informed route to learning compact effective response
models from limited finite-shot measurements generated by transmon-inspired
qutrit simulators. The two-qubit proof of concept shows that strongly incomplete
local data can still constrain a useful effective response representation within
the sampled hardware prior. The three-qubit extension gives a more cautious and
more informative scaling picture: local structured learning remains effective,
Ridge regression becomes the strongest local estimator in the reported scaled
runs, and targeted pair probes improve the prediction of process-relative
edge-level residuals.

The QAOA/MaxCut benchmark shows that learned response models can reproduce and
mitigate simulated cost-landscape distortions. Importantly, we distinguish
oracle-assisted diagnostics from non-oracle mitigation. The oracle diagnostic is
useful for testing whether the learned response model captures the simulated
hardware deformation, but it is not an experimental protocol. The non-oracle CDR-style correction avoids inserting the unknown ideal target
landscape and reduces the finite-shot QAOA landscape error from
$0.18187\,[0.17576,0.18798]$ to $0.01811\,[0.01751,0.01871]$ in the best
local-inverse variant.

The main conclusion is that hardware-informed measurement design, compact
local-plus-edge response models, leakage-explicit diagnostics, strong classical
baselines, and non-oracle regression-style mitigation form a promising
calibration strategy for near-term variational algorithms. In the scaled three-qubit local task, Ridge regression
provides the strongest baseline.

\section*{Data and code availability}
The simulation code needed to reproduce the reported numerical results is available at:
\url{https://github.com/ebrahimkh/transmon-error-learning-qaoa}.

\section*{Acknowledgments}

AI-assisted tools were used during the preparation of this manuscript. In particular, OpenAI ChatGPT (GPT-5.5 Thinking) was used for editorial refinement, including improving grammar, wording, clarity, and figure captions, and for code-organization suggestions. The scientific conception, simulation design, numerical experiments, code implementation, validation, analysis, interpretation, and conclusions were developed and verified by the authors, who take full responsibility for the work.


\onecolumngrid
\clearpage
\appendix

\section{Implementation details and hyperparameters}
\label{app:implementation_details}

This appendix summarizes the implemented simulator, datasets, machine-learning models, and numerical workflow used to generate the reported results.  The simulator parameters are hidden from the learner.  They are used only to generate finite-shot measurements and reference effective labels.

\subsection{Hidden Hamiltonian, control, and noise parameters}
\label{app:hamiltonian_parameters}

All Hamiltonian frequencies in the code are angular frequencies in rad/ns.  For readability, Table~\ref{tab:app_hamiltonian_params} reports frequency distributions in GHz before multiplication by $2\pi$.  Decoherence times are in ns.  The two-qubit and three-qubit simulators use the same physical scale, with the three-qubit model extending the topology to a chain.

\begin{table}[!htbp]
\caption{Hidden transmon-device parameter distributions used by the simulator.  Normal distributions are written as $\mathcal{N}(\mu,\sigma)$ and uniform distributions as $\mathcal{U}(a,b)$.  The learner never receives these microscopic parameters.}
\label{tab:app_hamiltonian_params}
\begin{ruledtabular}
\begin{tabular}{llll}
Quantity & Two-transmon implementation & Three-transmon implementation & Explanation \\
\hline
Hilbert truncation & $d=3$ per transmon & $d=3$ per transmon & Qutrit model with leakage level $\ket{2}$ \\
Topology & edge $(1,2)$ & chain edges $(1,2),(2,3)$ & Coupling graph \\
$\Delta_1/2\pi$ & $\mathcal{N}(0.000,0.004)$ GHz & $\mathcal{N}(0.000,0.006)$ GHz & Detuning/calibration offset \\
$\Delta_2/2\pi$ & $\mathcal{N}(0.150,0.006)$ GHz & $\mathcal{N}(0.080,0.006)$ GHz & Detuning/calibration offset \\
$\Delta_3/2\pi$ & -- & $\mathcal{N}(0.160,0.006)$ GHz & Detuning/calibration offset \\
$\alpha_1/2\pi$ & $\mathcal{N}(0.220,0.008)$ GHz & $\mathcal{N}(0.220,0.008)$ GHz & Anharmonicity magnitude \\
$\alpha_2/2\pi$ & $\mathcal{N}(0.230,0.008)$ GHz & $\mathcal{N}(0.224,0.008)$ GHz & Anharmonicity magnitude \\
$\alpha_3/2\pi$ & -- & $\mathcal{N}(0.228,0.008)$ GHz & Anharmonicity magnitude \\
$g_{ij}/2\pi$ & $\mathcal{N}(0.0030,0.0005)$ GHz & $\mathcal{N}(0.0030,0.0005)$ GHz per edge & Exchange-like coupling \\
$\zeta_{ij}/2\pi$ & $\mathcal{N}(0.0008,0.0002)$ GHz & $\mathcal{N}(0.0008,0.0002)$ GHz per edge & Residual $ZZ$-type shift \\
$s_{\Omega,i}$ & $\mathcal{N}(1.00,0.025)$ & $\mathcal{N}(1.00,0.025)$ & Hidden pulse-amplitude scale \\
$\phi_i$ & $\mathcal{N}(0,0.025)$ rad & $\mathcal{N}(0,0.025)$ rad & Hidden control-phase error \\
$T_{1,i}$ & $\mathcal{U}(25000,60000)$ ns & $\mathcal{U}(25000,60000)$ ns & Energy relaxation time \\
$T_{\phi,i}^{\rm coll}$ & $\mathcal{U}(20000,50000)$ ns & $\mathcal{U}(20000,50000)$ ns & Pure dephasing time \\
$r_{01,i}$ & $\mathcal{U}(0.005,0.040)$ & $\mathcal{U}(0.005,0.040)$ & Readout assignment error $\ket{0}\mapsto\ket{1}$ \\
$r_{10,i}$ & $\mathcal{U}(0.005,0.050)$ & $\mathcal{U}(0.005,0.050)$ & Readout assignment error $\ket{1}\mapsto\ket{0}$ \\
Pulse angle & $\theta=\pi/2$ & $\theta=\pi/2$ & Tomography/control pulse rotation \\
Pulse duration & $T=40$ ns & $T=40$ ns & Square-pulse duration \\
Pulse phase & $\phi_{\rm nominal}=0$ & $\phi_{\rm nominal}=0$ & Nominal control phase \\
Shot choices & $\{128,512,1024,4096\}$ & $\{128,512,1024,4096\}$ & Finite-shot tomography noise \\
QAOA shots & -- & $2048$ & Finite-shot QAOA landscape sampling \\
QAOA grid & -- & $31\times31$ grid & $\gamma\in[0,\pi]$, $\beta\in[0,\pi/2]$ \\
\end{tabular}
\end{ruledtabular}
\end{table}

\begin{figure*}[t]
\centering
\includegraphics[width=0.95\textwidth]{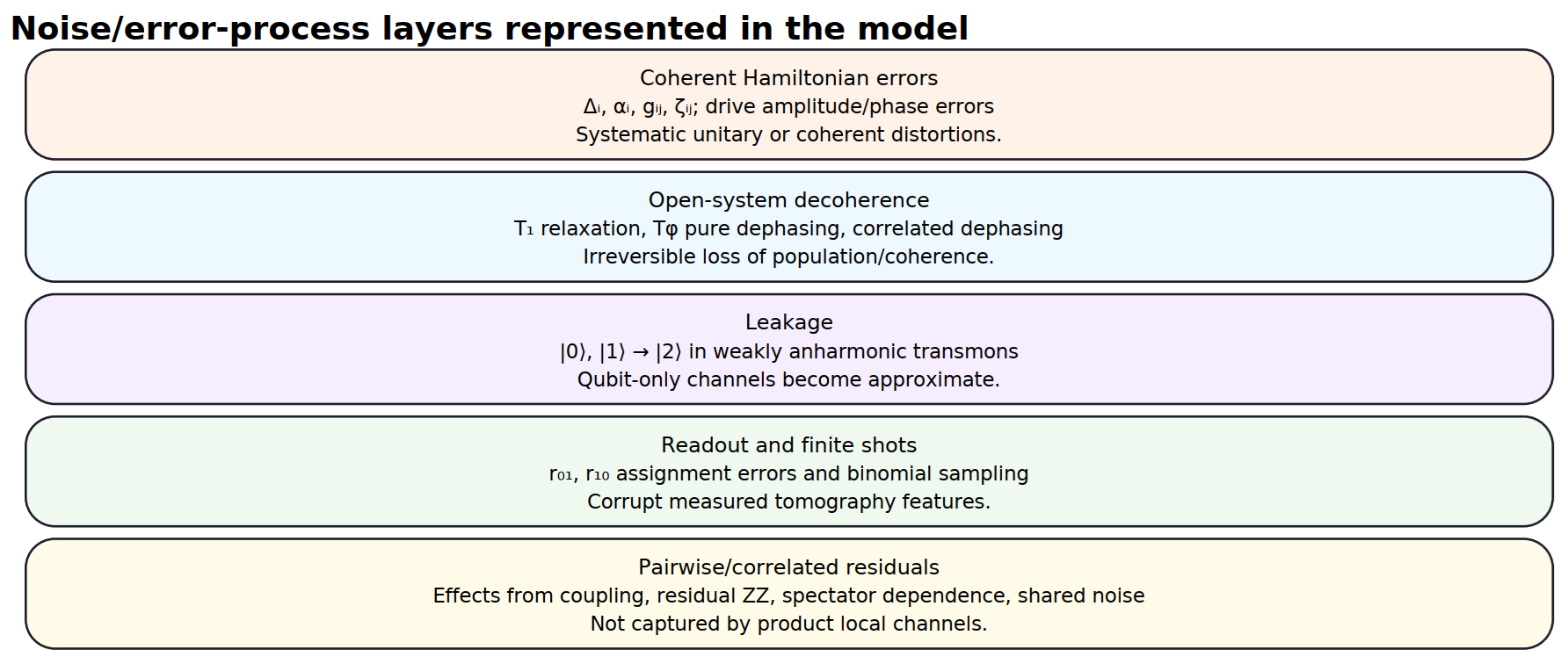}
\caption{Noise and error-process taxonomy.  The simulator explicitly includes coherent Hamiltonian imperfections, open-system decoherence, leakage, readout assignment error, finite-shot noise, and pairwise/correlated residual effects.}
\label{fig:noise_taxonomy}
\end{figure*}

The drift Hamiltonian is implemented as
\begin{equation}
H_0 =
\sum_i
\left[
\Delta_i n_i-\frac{\alpha_i}{2}n_i(n_i-1)
\right]
+
\sum_{(i,j)}
g_{ij}\left(a_i^\dagger a_j+a_i a_j^\dagger\right)
+
\sum_{(i,j)}
\zeta_{ij}n_i n_j ,
\end{equation}
with all single-transmon operators embedded in the full tensor-product Hilbert space.  The control Hamiltonian is implemented as a square microwave pulse.  For a target rotation angle $\theta$ and pulse duration $T$, the nominal Rabi scale is $\Omega=\theta/T$.  Hidden amplitude and phase errors modify this to
\begin{equation}
\widetilde{\Omega}_i=s_{\Omega,i}\Omega,\qquad
\widetilde{\phi}_i=\phi_{\rm nominal}+\phi_i ,
\end{equation}
and the control Hamiltonian on the driven transmon is
\begin{equation}
H_{\rm ctrl}^{(i)}(t)=
\frac{\widetilde{\Omega}_i\cos \widetilde{\phi}_i}{2}
\left(a_i+a_i^\dagger\right)
+
\frac{\widetilde{\Omega}_i\sin \widetilde{\phi}_i}{2}
i\left(a_i^\dagger-a_i\right).
\end{equation}
The total Hamiltonian used in the Lindblad evolution is $H(t)=H_0+H_{\rm ctrl}(t)$.

\subsection{Numerical simulation workflow}

Each dataset example is generated from a newly sampled hidden device.  The simulation workflow is:

\begin{enumerate}
\item Sample hidden transmon parameters from Table~\ref{tab:app_hamiltonian_params}.
\item Build qutrit operators $a_i$, $a_i^\dagger$, $n_i$, logical projectors, and embedded Pauli observables.
\item Construct $H_0$, $H_{\rm ctrl}(t)$, and the Lindblad collapse operators
\begin{equation}
L_{1,i}=\sqrt{1/T_{1,i}}\,a_i,\qquad
L_{\phi,i}=\sqrt{1/T_{\phi,i}^{\rm coll}}\,n_i .
\end{equation}
\item Evolve density matrices using a vectorized Liouvillian superoperator.  For a pulse duration $T$, the open-system propagator is $S=\exp(\mathcal{L}T)$.
\item Generate full finite-shot local tomography values from the hidden device.
\item Mask a subset of tomography entries to create the learner-visible measurement vector.
\item Compute the reference labels from the hidden simulator: local affine channels and, for the pair study, pair residuals.
\item Standardize features using training-set means and standard deviations.
\item Train supervised models to predict effective labels from limited measurements.
\item Insert the predicted effective error model into the QAOA landscape correction rule and compute reliability metrics.
\end{enumerate}

The learner therefore solves an operational inverse problem.  It receives finite-shot tomography features and the shot-count feature, but it never receives $\Delta_i$, $\alpha_i$, $g_{ij}$, $\zeta_{ij}$, $T_1$, $T_{\phi,i}^{\rm coll}$, readout errors, or pulse-calibration errors.

\subsection{Datasets and feature dimensions}

Table~\ref{tab:app_dataset_dimensions} summarizes the dataset dimensions.  In the three-qubit random-mask experiments, the masked local input has dimension
\begin{equation}
54 \; \text{values} + 54 \; \text{mask bits} + 1 \; \text{shot feature}=109 .
\end{equation}
For the pair-probe experiment, this is augmented by 18 pair-probe values and 18 pair-probe mask bits, giving input dimension $145$.

\begin{table}[!htbp]
\caption{Implemented dataset sizes, inputs, and labels.}
\label{tab:app_dataset_dimensions}
\begin{ruledtabular}
\begin{tabular}{lllll}
Experiment & Train/val/test & Input dimension & Output dimension & Measurement budget \\
\hline
Two-qubit partial tomography & $160/50/50$ & $13$ & $24$ & 12 local values + $\log_{10}N_{\rm shots}$ \\
Three-qubit local sweep & $500/100/100$ & $109$ & $36$ & $K=6,12,18,24,36,54$ local settings \\
Three-qubit pair-probe sweep & 400/80/80 & 145 & 198 & $K=18$ local settings, $K_{\rm pair}=0,6,12,18$ \\
Three-qubit QAOA evaluation & 50 devices & -- & -- & $K=18,54$ for local; all $K_{\rm pair}$ for pair study \\
\end{tabular}
\end{ruledtabular}
\end{table}

For the two-qubit partial tomography study, the observed local states are $\ket{0}$, $\ket{+}$, and $\ket{+i}$, and the measured Pauli bases are $X$ and $Z$.  This gives
\begin{equation}
2\;\text{qubits}\times 3\;\text{input states}\times 2\;\text{measurement bases}=12
\end{equation}
finite-shot tomography values, plus one shot-count feature.

For the three-qubit local study, full local tomography would contain
\begin{equation}
3\;\text{qubits}\times 6\;\text{input states}\times 3\;\text{measurement bases}=54
\end{equation}
values.  The random-mask protocol exposes only $K$ of these values to the learner.  The local label contains one affine map $(A_i,\bm b_i)$ per qubit, giving $3\times(9+3)=36$ output components.

For the pair-probe study, the physical edges are $(1,2)$ and $(2,3)$. Each edge
has the nine Pauli-pair probe settings $XX,XY,XZ,YX,YY,YZ,ZX,ZY,ZZ$.  The
learner-visible pair-probe input contains 18 pair-probe values and 18 pair-probe
mask bits.  The output label, however, is process-relative: each edge has nine
Pauli-pair input probes and nine Pauli-pair output responses, giving
$2\times 9\times 9=162$ pair-residual labels.  Together with the 36 local affine
channel labels, the augmented output dimension is $36+162=198$.

\subsection{Learning models and hyperparameters}

The following list summarizes the implemented models and hyperparameters used in the final reported run.

\begin{table*}[!htbp]
\caption{Implemented learning models and hyperparameters.}
\label{tab:implemented_models_hyperparams}
\footnotesize
\begin{description}[leftmargin=1.6em,itemsep=2pt,topsep=3pt]
\item[Mean baseline] Predicts the training-set mean label.  There are no fitted input-dependent parameters.
\item[Direct reconstruction] Uses a regularized least-squares update from observed tomography with ridge parameter $10^{-6}$; starts from the training-set mean label and updates only components constrained by the observed tomography entries.
\item[Ridge regression] Linear multi-output regression with $\ell_2$ regularization, regularization strength $\alpha=10^{-2}$, and standardized input features.
\item[Random Forest] Multi-output random forest with 120 trees, minimum samples per leaf set as in the notebook, and fixed random seed.
\item[Two-qubit ChannelNet] MLP architecture $13\rightarrow128\rightarrow128\rightarrow64\rightarrow24$; ReLU activations; AdamW optimizer; MSE loss on the 24-dimensional local-response label.
\item[Three-qubit ChannelNet] MLP architecture $109\rightarrow256\rightarrow256\rightarrow128\rightarrow36$; ReLU activations; AdamW optimizer; MSE loss on the 36-dimensional local-response label.  The input dimension has been corrected to match the $54$ values, $54$ mask bits, and one shot-count feature stated in Table~\ref{tab:app_dataset_dimensions}.
\item[PairProbeAwareNet] Shared trunk $145 \rightarrow 320 \rightarrow 256 \rightarrow 128$,
channel head $128 \rightarrow 36$, and pair head $128 \rightarrow 96 \rightarrow 162$;
ReLU activations; AdamW optimizer; weighted multitask MSE loss for local-response
and process-relative pair-residual labels.
\end{description}
\end{table*}

All neural networks use fully connected layers with ReLU activations after each hidden layer and a linear output layer.  Inputs and labels are standardized using the training-set mean and standard deviation.  The validation set is used to select the best model state according to validation mean-squared error or, in the pair-probe case, a combined validation channel-plus-pair score.

The two-qubit and three-qubit ChannelNet losses have the form
\begin{equation}
\mathcal{L}_{\rm local}
=
\mathrm{MSE}(\hat{\bm y}_{\rm ch},\bm y_{\rm ch})
+
\lambda_{\rm phys}\mathcal{L}_{\rm phys},
\qquad
\lambda_{\rm phys}=0.03 .
\end{equation}
The physicality penalty evaluates the predicted affine maps on a fixed set of test Bloch vectors and penalizes predicted output vectors whose norm exceeds one.  This softly discourages unphysical affine maps without imposing a full complete-positivity constraint.

The pair-probe-aware model uses the multitask loss
\begin{equation}
\mathcal{L}_{\rm pair}
=
\mathrm{MSE}(\hat{\bm y}_{\rm ch},\bm y_{\rm ch})
+
\lambda_{\rm pair}\mathrm{MSE}(\hat{\bm\delta},\bm\delta)
+
\lambda_{\rm phys}\mathcal{L}_{\rm phys},
\end{equation}
with $\lambda_{\rm pair}=1.0$ and $\lambda_{\rm phys}=0.03$.  The output vector has 198 components: the first 36 are the local affine-channel
labels and the final 162 are process-relative pair residuals.

\subsection{Classical baselines}

The direct reconstruction baseline uses the observed tomography settings to solve small regularized linear systems for the measured rows of the affine map.  Unobserved components are initialized from the training-set mean channel. The direct baseline is strong when the tomography budget is high, but weak when the measurement mask is sparse.  In the main text, conclusions about neural-network advantage should therefore be restricted to the underdetermined two-qubit case; the scaled three-qubit data support the stronger and more conservative claim that structured linear baselines are highly competitive.

Ridge regression is trained on standardized masked-tomography inputs.  It is an important baseline because the effective label is itself an affine input--output representation; consequently, part of the inverse problem is expected to be close to linear.  Random Forest regression is included as a nonlinear non-neural baseline.  Its weaker performance in the final three-qubit scaled run indicates that the chosen local representation is better matched to linear or smooth function approximators than to the tree ensemble used here.

\subsection{QAOA evaluation implementation}

The QAOA benchmark uses depth $p=1$.  The three-qubit graph is the chain with edges $(1,2)$ and $(2,3)$.  The ideal QAOA landscape is evaluated on a $31\times31$ grid with
\begin{equation}
\gamma\in[0,\pi],\qquad \beta\in[0,\pi/2].
\end{equation}
For the oracle-diagnostic evaluation, the code computes the ideal landscape, the noisy landscape, the model-predicted noisy landscape, and the oracle-diagnostic bias-corrected landscape,
\begin{equation}
C_{\rm mit}(\gamma,\beta)
=
C_{\rm noisy}(\gamma,\beta)
-
\left[
C_{\rm model}(\gamma,\beta)-C_{\rm ideal}(\gamma,\beta)
\right].
\end{equation}
This expression is used only for the oracle-assisted simulation diagnostic. For
the non-oracle evaluation, the CDR local-inverse protocol trains a
regression map on auxiliary classically tractable circuits and applies the
learned map to the target noisy QAOA landscape without inserting the ideal target
landscape.
The reported metrics are then computed from this grid: landscape MAE, RMSE, selected-point regret, approximation ratio, optimal-bitstring probability, and parameter displacement.
In this implementation, \(C_{\rm noisy}\) is obtained by applying the
reference effective response model, computed from the hidden qutrit simulator,
to the ideal logical QAOA density matrices and then sampling finite shots.
It is not obtained from a full qutrit pulse-level simulation of the complete
QAOA circuit.

\subsection{Leakage-aware diagnostic}

The simulator also records leakage labels in addition to logical affine
response labels.  This allows us to test whether explicitly learning leakage
diagnostics improves the effective response model.  We compare a channel-only
model with a leakage-aware multitask model that predicts both the local affine
channel and leakage-related labels.

Table~\ref{tab:leakage_aware_diagnostic} summarizes the leakage-aware diagnostic.
The leakage-aware multitask model reduces the channel MSE from
$7.511\times 10^{-2}$ to $6.225\times 10^{-2}$ and reduces the mean channel L2
error from $1.5863$ to $1.4653$.  It also predicts leakage with leakage MAE
$2.159\times 10^{-2}$.  In the corresponding QAOA diagnostic, the channel-only
model reduces the landscape MAE from $0.15879$ to $0.04406$, while the
leakage-aware model further reduces it to $0.03944$.  The improvement ratio
therefore increases from $3.60\times$ to $4.03\times$.  This confirms that
leakage-explicit labels are useful as diagnostics, even though the main learned
response model remains a compact logical affine representation supplemented by
process-relative edge residuals.

\begin{table}[t]
\centering
\caption{
Leakage-aware diagnostic for the three-qubit local response model.  The
leakage-aware multitask model predicts both local affine-channel labels and
leakage diagnostics.  The QAOA improvement is computed relative to the noisy
baseline in the same leakage-aware test set.
}
\begin{tabular}{lcccc}
\hline\hline
Model & Channel MSE & Channel L2 & QAOA MAE & Improvement \\
\hline
Noisy baseline & -- & -- & $0.15879$ & $1.00\times$ \\
Channel-only & $7.511\times 10^{-2}$ & $1.5863$ & $0.04406$ & $3.60\times$ \\
Leakage-aware multitask & $6.225\times 10^{-2}$ & $1.4653$ & $0.03944$ & $4.03\times$ \\
\hline\hline
\end{tabular}
\label{tab:leakage_aware_diagnostic}
\end{table}

\renewcommand{\topfraction}{0.95}
\renewcommand{\bottomfraction}{0.90}
\renewcommand{\textfraction}{0.05}
\renewcommand{\floatpagefraction}{0.85}

\setcounter{topnumber}{8}
\setcounter{bottomnumber}{8}
\setcounter{totalnumber}{12}


\begin{thebibliography}{99}
\bibitem{Koch2007Transmon} J. Koch, T. M. Yu, J. Gambetta, A. A. Houck, D. I. Schuster, J. Majer, A. Blais, M. H. Devoret, S. M. Girvin, and R. J. Schoelkopf, Charge-insensitive qubit design derived from the Cooper pair box, Phys. Rev. A \textbf{76}, 042319 (2007).
\bibitem{Krantz2019Guide} P. Krantz, M. Kjaergaard, F. Yan, T. P. Orlando, S. Gustavsson, and W. D. Oliver, A quantum engineer's guide to superconducting qubits, Appl. Phys. Rev. \textbf{6}, 021318 (2019).
\bibitem{Blais2021CircuitQED} A. Blais, A. L. Grimsmo, S. M. Girvin, and A. Wallraff, Circuit quantum electrodynamics, Rev. Mod. Phys. \textbf{93}, 025005 (2021).
\bibitem{Motzoi2009DRAG} F. Motzoi, J. M. Gambetta, P. Rebentrost, and F. K. Wilhelm, Simple pulses for elimination of leakage in weakly nonlinear qubits, Phys. Rev. Lett. \textbf{103}, 110501 (2009).
\bibitem{Lindblad1976} G. Lindblad, On the generators of quantum dynamical semigroups, Commun. Math. Phys. \textbf{48}, 119--130 (1976).
\bibitem{Gorini1976} V. Gorini, A. Kossakowski, and E. C. G. Sudarshan, Completely positive dynamical semigroups of N-level systems, J. Math. Phys. \textbf{17}, 821--825 (1976).
\bibitem{Sarovar2020Crosstalk} M. Sarovar, T. Proctor, K. Rudinger, K. Young, E. Nielsen, and R. Blume-Kohout, Detecting crosstalk errors in quantum information processors, Quantum \textbf{4}, 321 (2020).
\bibitem{Ni2022ResidualZZ} Z. Ni, S. Li, L. Zhang, et al., Scalable method for eliminating residual ZZ interaction between superconducting qubits, Phys. Rev. Lett. \textbf{129}, 040502 (2022).
\bibitem{Rudinger2021SimGST} K. Rudinger, C. W. Hogle, R. K. Naik, A. Hashim, D. Lobser, D. I. Santiago, M. D. Grace, E. Nielsen, T. Proctor, S. Seritan, S. M. Clark, R. Blume-Kohout, I. Siddiqi, and K. C. Young, Experimental characterization of crosstalk errors with simultaneous gate set tomography, PRX Quantum \textbf{2}, 040338 (2021).
\bibitem{Nielsen2021GST} E. Nielsen, J. K. Gamble, K. Rudinger, T. Scholten, K. Young, and R. Blume-Kohout, Gate set tomography, Quantum \textbf{5}, 557 (2021).
\bibitem{Gross2010Compressed} D. Gross, Y.-K. Liu, S. T. Flammia, S. Becker, and J. Eisert, Quantum state tomography via compressed sensing, Phys. Rev. Lett. \textbf{105}, 150401 (2010).
\bibitem{Rodionov2014CompressedQPT} A. V. Rodionov, A. Veitia, R. Barends, et al., Compressed sensing quantum process tomography for superconducting quantum gates, Phys. Rev. B \textbf{90}, 144504 (2014).
\bibitem{Torlai2023TensorQPT} G. Torlai, C. J. Wood, A. Acharya, G. Carleo, J. Carrasquilla, and L. Aolita, Quantum process tomography with unsupervised learning and tensor networks, Nat. Commun. \textbf{14}, 2858 (2023).
\bibitem{Gaikwad2024NNTomography} A. Gaikwad, O. Bihani, Arvind, and K. Dorai, Neural-network-assisted quantum state and process tomography using limited data sets, Phys. Rev. A \textbf{109}, 012402 (2024).
\bibitem{Ahmed2023KrausLearning} S. Ahmed, F. Quijandria, and A. F. Kockum, Gradient-descent quantum process tomography by learning Kraus operators, Phys. Rev. Lett. \textbf{130}, 150402 (2023).
\bibitem{Temme2017QEM} K. Temme, S. Bravyi, and J. M. Gambetta, Error mitigation for short-depth quantum circuits, Phys. Rev. Lett. \textbf{119}, 180509 (2017).
\bibitem{Endo2018PracticalQEM} S. Endo, S. C. Benjamin, and Y. Li, Practical quantum error mitigation for near-future applications, Phys. Rev. X \textbf{8}, 031027 (2018).
\bibitem{Cai2023QEMReview} Z. Cai, R. Babbush, S. C. Benjamin, S. Endo, W. J. Huggins, Y. Li, J. R. McClean, and T. E. O'Brien, Quantum error mitigation, Rev. Mod. Phys. \textbf{95}, 045005 (2023).
\bibitem{Strikis2021LearningQEM} A. Strikis, D. Qin, Y. Chen, S. C. Benjamin, and Y. Li, Learning-based quantum error mitigation, PRX Quantum \textbf{2}, 040330 (2021).
\bibitem{Czarnik2021CDR} P. Czarnik, A. Arrasmith, P. J. Coles, and L. Cincio, Error mitigation with Clifford quantum-circuit data, Quantum \textbf{5}, 592 (2021).
\bibitem{Liao2024MLQEM} H. Liao, D. S. Wang, I. Sitdikov, C. Salcedo, A. Seif, and Z. K. Minev, Machine learning for practical quantum error mitigation, Nat. Mach. Intell. \textbf{6}, 1478--1486 (2024).
\bibitem{Farhi2014QAOA} E. Farhi, J. Goldstone, and S. Gutmann, A quantum approximate optimization algorithm, arXiv:1411.4028 (2014).
\bibitem{Blekos2024QAOAReview} K. Blekos, D. Brand, A. Ceschini, C.-H. Chou, R.-H. Li, K. Pandya, and A. Summer, A review on Quantum Approximate Optimization Algorithm and its variants, Phys. Rep. \textbf{1068}, 1--66 (2024).
\bibitem{Zhou2020QAOA} L. Zhou, S.-T. Wang, S. Choi, H. Pichler, and M. D. Lukin, Quantum approximate optimization algorithm: Performance, mechanism, and implementation on near-term devices, Phys. Rev. X \textbf{10}, 021067 (2020).
\bibitem{Lange2023ActiveQST} H. Lange, M. Kebric, M. Buser, U. Schollwoeck, F. Grusdt, and A. Bohrdt, Adaptive quantum state tomography with active learning, Quantum \textbf{7}, 1129 (2023).
\bibitem{Quek2021AdaptiveNN} Y. Quek, S. Fort, and H. K. Ng, Adaptive quantum state tomography with neural networks, npj Quantum Inf. \textbf{7}, 105 (2021).
\bibitem{Yang2025ActiveQPT} J. Yang, X. Xu, and W. Xie, Active learning with variational quantum circuits for quantum process tomography, arXiv:2412.20925 (2024).
\bibitem{NielsenChuang} M. A. Nielsen and I. L. Chuang, \textit{Quantum Computation and Quantum Information}, 10th anniversary ed. (Cambridge University Press, Cambridge, 2010).

\bibitem{KingRuskai2001} C. King and M. B. Ruskai, Minimal entropy of states emerging from noisy quantum channels, IEEE Trans. Inf. Theory \textbf{47}, 192--209 (2001).

\bibitem{Ruskai2002} M. B. Ruskai, S. Szarek, and E. Werner, An analysis of completely-positive trace-preserving maps on $M_2$, Linear Algebra Appl. \textbf{347}, 159--187 (2002).

\bibitem{WoodGambetta2018} C. J. Wood and J. M. Gambetta, Quantification and characterization of leakage errors, Phys. Rev. A \textbf{97}, 032306 (2018).

\bibitem{WillmottMatsuura2005} C. J. Willmott and K. Matsuura, Advantages of the mean absolute error (MAE) over the root mean square error (RMSE) in assessing average model performance, Clim. Res. \textbf{30}, 79--82 (2005).

\bibitem{GoemansWilliamson1995} M. X. Goemans and D. P. Williamson, Improved approximation algorithms for maximum cut and satisfiability problems using semidefinite programming, J. ACM \textbf{42}, 1115--1145 (1995).

\bibitem{Moll2018Variational} N. Moll, P. Barkoutsos, L. S. Bishop, J. M. Chow, A. Cross, D. J. Egger, S. Filipp, A. Fuhrer, J. M. Gambetta, M. Ganzhorn, A. Kandala, A. Mezzacapo, P. M\"uller, W. Riess, G. Salis, J. Smolin, I. Tavernelli, and K. Temme, Quantum optimization using variational algorithms on near-term quantum devices, Quantum Sci. Technol. \textbf{3}, 030503 (2018).
\end{thebibliography}
\end{document}